\documentclass[twocolumn,showpacs,preprintnumbers,amsmath,amssymb]{revtex4}

\usepackage{graphicx}% Include figure files
\usepackage{dcolumn}% Align table columns on decimal point
\usepackage{bm}% bold math
\newcommand{\ds}{\displaystyle{}}
\newcommand{\fsm}[6]{\displaystyle{}\frac{#1}{#2-\displaystyle{}\frac{#3}{#4-\displaystyle{}\frac{#5}{#6-\ddots}}}}

\begin{document}

\preprint{APS/123-QED}

\title{Stark effect upon the effective mass and radius \\ in a tight-binding exciton model.}% Force line breaks with \\

\author{Jean El-khoury}
\email{elkhoury@ucla.edu}

\author{Jean-Pierre Gallinar}%
\email{jpgal@usb.ve}
\affiliation{%
Departamento de F\'{\i}sica, Universidad Sim\'on Bol\'{\i}var,\\
Apartado 89000, Caracas 1080A, Venezuela}

\date{\today}% It is always \today, today,
             % but any date may be explicitly specified

\begin{abstract}
 With a Green's function formalism we obtain the eigenvalue
 spectrum of a tight-binding one-dimensional exciton model
 characterized by a contact interaction, a Coulombic
 electron and hole attraction, the Heller-Marcus
 exciton-hopping energy and an external constant and homogeneous
 electric field. The resulting eigenvalue spectrum, in the form
 of an unevenly spaced Wannier-Stark ladder with envelope
 profiles, is used to obtain the effective mass of the
 exciton by the application of the Mattis-Gallinar effective
 mass formula [D. C. Mattis and J.-P. Gallinar,
 {\it Phys. Rev. Lett.} {\bf{53}}, 1391 (1984)]. We obtain positive
 and negative effective masses for the exciton. The inverse
 effective mass may oscillate periodically as a function of
 the inverse of the electric field, with the frequency of
 oscillation linearly dependent upon the tight-binding hopping
 matrix element. The exciton radius is also obtained with the
 Green's function formalism, and it too exhibits Keldysh-like
 field dependent oscillations, as well as abrupt variations
 associated to strongly avoided crossings in the eigenvalue spectrum.
 Finally, some comments are made about the experimental relevance
 of our results.
\end{abstract}

% PACS, the Physics and Astronomy Classification Scheme.

\maketitle

\section{\label{sec:intro}Introduction}
As shown by Mattis and Gallinar \cite{MG-1984,Mattis}, the mass of
an exciton is not, in general, given by the intuitive result of
the sum of its electron and hole masses, but it may depend upon
excitonic internal quantum numbers, such as, for example, the
internal kinetic energy of the exciton \cite{MG-1984}, or the
Heller-Marcus exciton-hopping energy \cite{GM-1985}. Beautiful
experimental confirmation of some of these surprising, and
interesting, predictions, was given by Cafolla et al.
\cite{cafolla} in $1985$ from electron-energy-loss spectroscopy.
Up to now, however, no detailed study or analysis of a specific
exciton model \cite{G-1996} has been made \footnote{With or
without an externally applied electric field.} in the light of the
Mattis-Gallinar formula (MGF) (cf. Eq.(\ref{eq:MGF})) and of its interesting
counterintuitive predictions. In view of this, and to further
study these predictions, here we shall present a specific, simple,
one-dimensional exciton model susceptible to an exact analysis for
its eigenvalue spectrum, even while in the presence of an external
electric field. Thus enabling the study (in an exact manner) of
diverse aspects related not only to the mass of an exciton through
the application of the (MGF), but also of those related, for
example, to another important general exciton property, such as
its radius. As we shall see, some of the shared characteristics of
these two exciton parameters (mass and radius) can be understood
in terms of the underlying eigenvalue spectrum of the exciton,
which, due to the applied external electric field, is obtained to
be in the form of an unevenly spaced Wannier-Stark ladder. This
ladder, in turn, exhibits Keldysh-like oscillations
\cite{callaway} and strongly avoided crossings of energy levels,
with dramatic consequences for the mass and radius of the exciton
worthy of close examination.

The plan of this paper is then as follows: in Section
\ref{sec:model} we present the mathematical Green's function
formalism to be used to obtain the eigenvalue spectrum, as well as
the different parameters that go into the exciton model to be
considered. In Section \ref{sec:results} we exhibit a collection
of graphical results, and in Section \ref{sec:conclusions}
conclusions are presented and discussed. Finally, an Appendix with
several analytical results is included.

\section{\label{sec:model}Exciton model}
The one-dimensional exciton model that we will consider
\cite{GM-1985}, akin to that of the Merrifield \cite{merrifield}
exciton, can be defined in terms of the matrix elements of the
excitonic Hamiltonian ${\cal H}$. If $\vert n,m \rangle$
represents a Wannier orthonormal basis states in which the
electron is localized at site $n$, and the hole at site $m$, the
nonvanishing matrix elements of ${\cal H}$ will be taken to be
\begin{equation}
\langle n,m \vert {\cal H} \vert n,m \rangle = W +V(n-m),
\label{eq:ec-1}
\end{equation}
where $W$ represents the combined half-bandwidth of the electron
plus hole, $n-m$ is the relative electron-hole coordinate and
$V(n-m)$ is the effective two-body potential through which the
electron and hole interact with each other \cite{merrifield}, and
with the external electric field, if present \footnote{No external
field is considered in \cite{merrifield}.}. We will assume that
$V(n-m)$ is given by \cite{merrifield}
\begin{equation}
V(n)= -V_o \, \delta_{n,0} -(1-\delta_{n,0}) \,\frac{V}{|n|}+
\alpha \, n,
\label{eq:ec-2}
\end{equation}
where $n=0,\pm 1, \pm 2,...$, $V_o$ measures the contact
interaction \cite{merrifield}, or electron and hole attraction
when they are at the same site on top of each other, while, on the
other hand, $V$ measures the long-range \cite{merrifield} part of
the attraction, modelled here by a Coulombic $1/|n|$ tail.
Finally, the interaction of the exciton electric dipole moment
with the external electric field $E$ is given by $\alpha\,n$, with
$\alpha = a\, e\, E$, where $`` a "$ is the lattice constant and
$(-e )$ the electron charge.

The non-diagonal matrix elements of ${\cal H}$ associated to the
kinetic energy of the exciton are given by
\begin{eqnarray}
\langle n , m \vert {\cal H} \vert n , m \pm 1 \rangle = -t_h ,\hskip0.4cm \mbox{and}
\nonumber
\\ \nonumber \\ \langle n , m \vert {\cal H} \vert n \pm 1 ,
m \rangle =-t_e .
\end{eqnarray}

With $t_h$ representing the usual valence-band width parameter
associated with nearest-neighbor jumping of the hole, and
$t_e$ the corresponding conduction-band width parameter
associated to the electron. Thus, $W = 2 (t_h +t_e)$.
Finally, we introduce the exciton-hopping (or Heller-Marcus)
matrix elements given by
\begin{equation}
\langle m , m \vert {\cal H} \vert n , n \rangle = H(n-m),
\hskip0.4cm n \ne m
\end{equation}
whereby the entire (Frenkel-like) exciton jumps as a whole entity
between sites $n$ and $m$, with the amplitude $H(n-m)$.

According to the (MGF) \cite{MG-1984,GM-1985}, the translational mass
$M_n$ of an exciton can be written as
\begin{equation}
\displaystyle{}M_n=\displaystyle{} \frac{m_e +
m_h}{1-\displaystyle{}\frac{K_n}{W}+\displaystyle{}
\frac{m_e+m_h}{M_F}\frac{H_n}{H_F}},
\label{eq:MGF}
\end{equation}
where $m_e$ and $m_h$ are, respectively, the electron and hole
effective masses, $K_n$ and $H_n$ are the expectation values of
the kinetic and exciton-hopping energies of the exciton in the
$nth$ bound or localized state of the interaction potential, and,
finally, $M_F$ and $H_F$ are the values respectively taken by
$M_n$ and $H_n$ for a Frenkel-like, strongly localized exciton.

As Eq (\ref{eq:MGF}) shows, the mass $M_n$ of the exciton can be
determined once the explicit values of $K_n$ and $H_n$ are given.
These values can be obtained through the Hellmann-Feynman theorem,
by writing \cite{GM-1985}
\begin{eqnarray}
K_n =&& \left(\frac{\partial E_n}{\partial
\lambda}\right)_{\lambda=\mu=1}, \,\,\,\,\, \,\mbox{and}\nonumber
\\ \nonumber \\ H_n =&&\left(\frac{\partial E_n}{\partial \mu}
\right)_{\lambda=\mu=1} ;
\label{eq:hellman}
\end{eqnarray}
where $E_n \equiv E_n(k=0)$ is the eigenvalue of ${\cal H}$ in the
$nth$ bound or localized state of the exciton at $k=0$, with $k$
being the center-of-mass wave vector of the exciton
\cite{GM-1985}. In Eq. (\ref{eq:hellman}), $E_n$ is made to depend
upon the dimensionless parameters $\lambda$ and $\mu$ through the
following simple scaling \cite{GM-1985}

\begin{eqnarray}
t_h \rightarrow {\lambda}\, t_h \nonumber \\ \nonumber \\ t_e
\rightarrow {\lambda}\, t_e,\nonumber
\end{eqnarray}
and
\begin{equation}
H(n-m) \rightarrow  \mu\, H(n-m).
\end{equation}

A Green's function formalism \cite{G-1984,economou} permits then the
evaluation \cite{GM-1985} of the energy eigenvalues $E_n$, by finding
the simple poles $z_n(m)$ of the continued fraction $G_{m,m}(z)$ given by
\begin{eqnarray}
G_{m,m}^{-1}(z) =&& z-V(m)-\frac{\varepsilon^2}{z-V(m+1)-\ddots}
\nonumber \\ \nonumber \\ && -
\frac{\varepsilon^2}{z-V(m-1)-\ddots}.
\label{eq:frac}
\end{eqnarray}
Where $\varepsilon= \lambda \,(t_e+t_h)$, and $m=0,\pm 1, \pm 2,
...$ is a fixed integer. Due to the peculiar nature of the
Heller-Marcus \footnote{The Heller-Marcus term acts only when the
electron and hole are at the same lattice site.} exciton-hopping
energy \cite{GM-1985}, the Term $V(0)$ appearing in Eq.
(\ref{eq:frac}) is shifted so that \cite{GM-1985} $V(0)
\rightarrow V(0) + \mu H_F$. The set of all eigenvalues $E_n$ of
${\cal H}$ can then be found and ordered appropriately
\cite{economou} from the relationships $E_n =z_n(m) + W$, in a
manner over-all independent of the chosen m's (from now on for
simplicity we write $z_n(m)\equiv z_n$).

The mean-square radius $R^2_n$ of the exciton can also be found in
terms of $G_{m,m}^{-1}(z)$. In fact, the mean-square radius of the
exciton is defined in terms of the relative electron-hole
coordinate $m$, as
\begin{equation}
R^2_n=\sum_{m=-\infty}^{m=+\infty}a^2\,m^2|F_n(m)|^2,
\label{eq:radiusf}
\end{equation}
where $F_n(m)$ is the normalized relative coordinate wavefunction
of the exciton \cite{GM-1985}. As shown in the Appendix, the
Green's function formalism \cite{economou} permits then the direct
evaluation of $|F_n(m)|^2$, obtaining finally
\begin{equation}
R_n^2=\sum_{m=-\infty}^{m=+\infty}\frac{a^2 m^2}{\left(
\frac{\partial}{\partial z} G_{m,m}^{-1}(z)\right)_{z=z_n}}.
\label{eq:radius}
\end{equation}

In contrast with the evaluation of the set of eigenvalues $E_n$,
all values of $m$ are seen in Eq (\ref{eq:radius}) to be in
principle relevant for the evaluation of $R^2_n$. In the next
section 3 we shall display some of the results obtained through
the use of the above described formalism applied to our exciton
model.
\section{\label{sec:results}Results}
In this section we display three types of results: energy spectra,
effective masses and root-mean-square radii of the exciton.
Figures (1) and (2) were obtained starting from the numerical
evaluation \cite{numericalrecipes} of the continued fraction in
Eq.(\ref{eq:frac}). In Fig (1), $m$ was chosen for evaluation
purposes as $m=0$ due to absence of Coulomb interaction, while in
Fig (2) $m$'s different from zero were appropriately chosen for
ease of computation. Because of the contact interaction $V_o$
being $V_o/\alpha=10$, level $n=0$ in the ladder in Fig (1) is
moved down to the level $n=-10$ where a degeneracy occurs at
$t=0$. While in Fig (2), with Coulomb interaction, no such
degeneracy occurs for the selected parameters. In both Figures (1)
and (2), several (strongly) avoided energy crossings are evident
and are associated to the envelope profiles. In Fig (1), the
envelope profile which starts at $z_n/\alpha=-10$ is that of the
single bound state \cite{13} of the contact interaction $V_o$ (in
absence of the electric field), i.e.,
\begin{equation}
E_n-W=-\sqrt{{V_o}^2+W^2}.
\label{eq:boundstate}
\end{equation}

%%%%%%%%%%%
\begin{figure}[htp]
\centering
\includegraphics[scale=0.38]{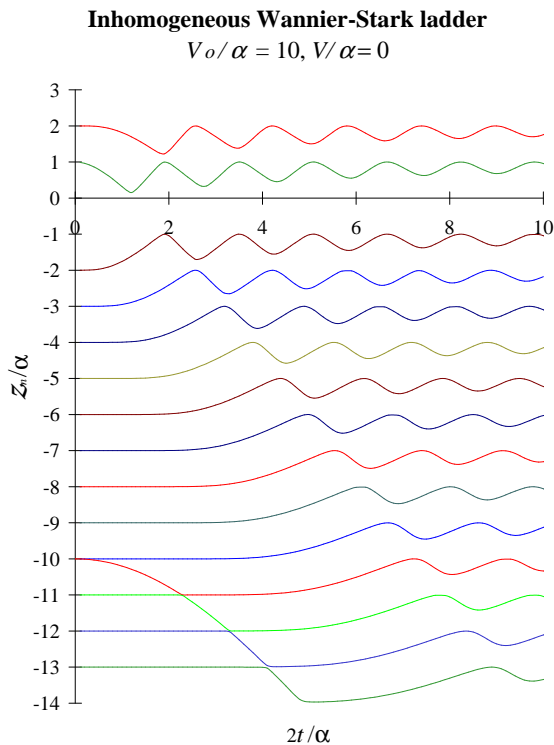}
\caption{Inhomogeneous Wannier-Stark Ladder energy levels $E_n -
W$ for the exciton in terms of z$_n(m)/\alpha$ vs $2\,t/\alpha$,
with $V=0$ and $V_o/\alpha=10$; $n$ runs from $-13$ to $2$ (skipping $n=-1$ for visual convenience),
the chosen $m$ is zero and henceforth $t \equiv t_e=t_h$. Note the increment in the degree
of flatness down in the ladder where a large number of derivatives is identically zero at $t=0$.
Noteworthy characteristics of the ladder are: oscillations as $t$ increases, strongly avoided energy
crossings and envelope profiles. Energy levels $n=0$ and $n=-10$ are degenerate at $t=0$.}
\label{fig:page61}
\end{figure}

\begin{figure}[htp]
\centering
\includegraphics[scale=0.45]{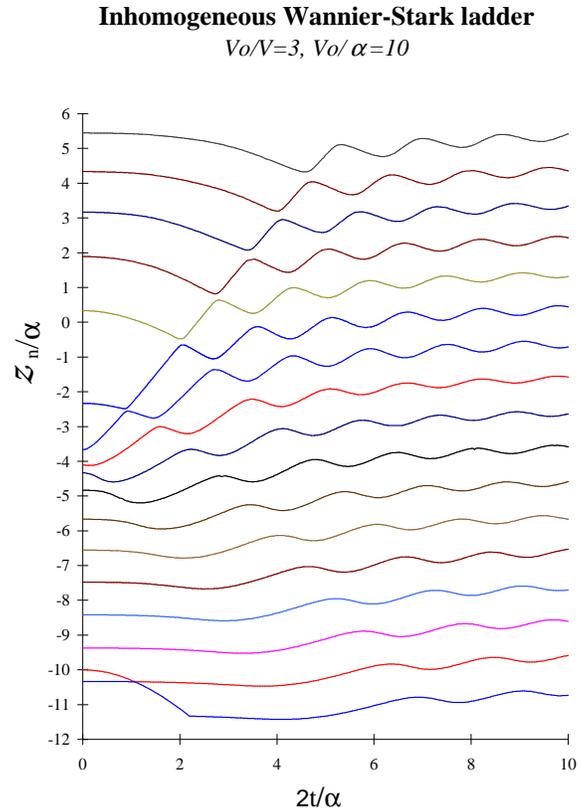}
\caption{Inhomogeneous Wannier-Stark Ladder energy levels $E_n -
W$ for the exciton in terms of z$_n(m)/\alpha$ vs $2\,t/\alpha$,
with $V_o/V=3$ and $V_o/\alpha=10$; $n$ runs from $-10$ to $6$,
the chosen $m$'s are not all zero and $t \equiv t_e=t_h$. Note the increment in the degree
of flatness down in the ladder where a large number of derivatives is vanishingly small at $t=0$.
Noteworthy characteristics of the ladder are: oscillations as $t$ increases, strongly avoided energy
crossings and envelope profiles. }
\label{fig:page20}
\end{figure}

\begin{figure}[htp]
\centering
\includegraphics[scale=0.4]{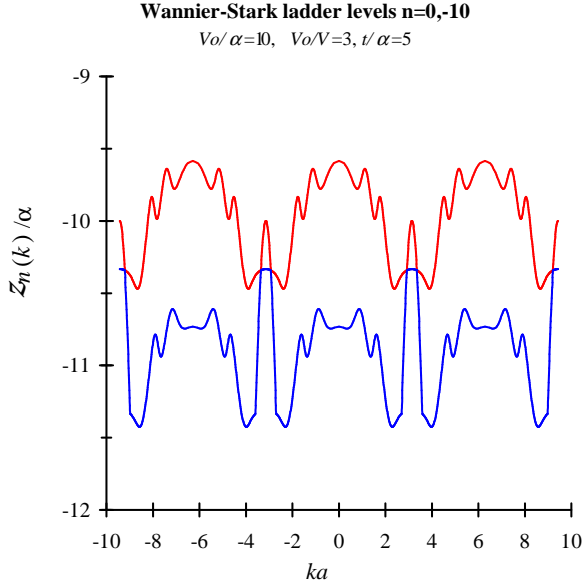}
\caption{Plot of Wannier-Stark ladder levels $n=0$ (red) and $n=-10$ (blue) vs $ka$
in the first three Brillouin Zones. The strongly avoided crossing of corresponding
levels shown in figure \ref{fig:page20} appears here close to the
borders of the Brillouin Zones.   }
\label{fig:brillouin}
\end{figure}

\begin{figure}[htp]
\centering
\includegraphics[width=8.5cm,height=7cm]{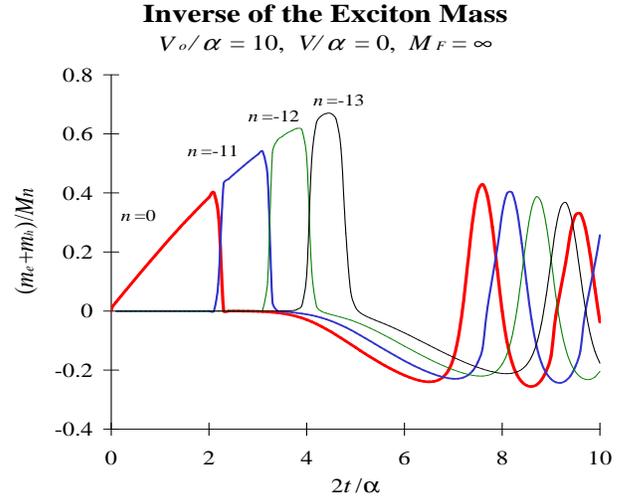}
\caption{Plot of the inverse effective dimensionless mass $(m_e+m_h)/M_n$ vs
$2\,t/\alpha$, with $V_o/\alpha=10$ and $M_{F}=\infty$, for the energy
levels $n=0,-11,-12,-13$. Both here and in the Wannier-Stark ladder in figure \ref{fig:page61},
these levels give rise to a noticeable visual envelope and abrupt changes in the exciton
mass. The amplitude of the oscillations decreases for large values of $2\,t/\alpha$.}\label{fig:page68}
\end{figure}

\begin{figure}[htp]
\centering
\includegraphics[width=8.5cm,height=7cm]{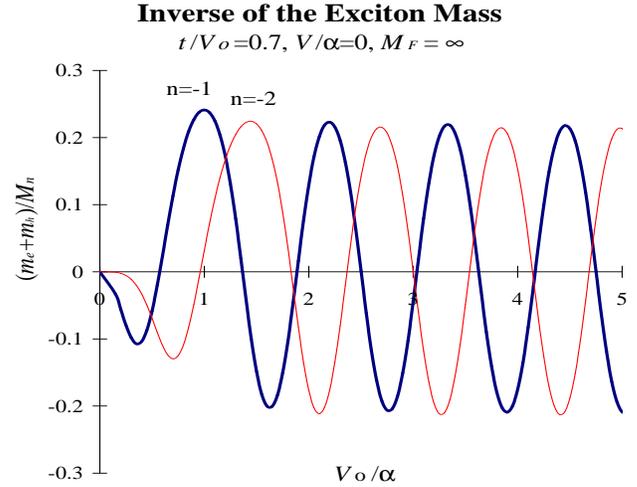}
\caption{Plot of the inverse effective dimensionless mass $(m_e+m_h)/M_n$ vs $V_o/\alpha$, with
$t/V_o=0.7$, $V/V_o=0$ and $M_{F}=\infty$, for two energy levels ($n = -1, -2$)
from the Wannier-Stark Ladder in figure \ref{fig:page61}. As $\alpha^{-1} \rightarrow \infty$,
the curves become periodic in the inverse electric field $1/\alpha$, with period $P$ given by
$P=\frac{\pi}{4}(V_o/t)\approx 1.122$ .}
\label{fig:page70}
\end{figure}

\begin{figure}[htp]
\centering
\includegraphics[width=8.5cm,height=7cm]{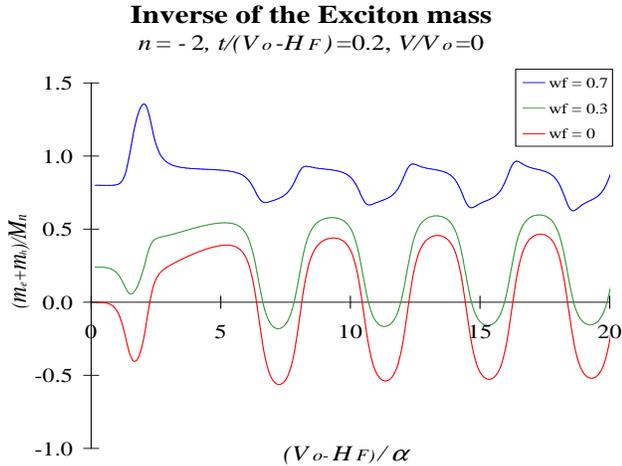}
\caption{Plot of the inverse effective dimensionless mass $(m_e+m_h)/M_n$ vs $V_o/\alpha$
for the energy level $n = -2$, with $t/V_o=0.2$, $V/V_o=0$, and three values of the
Wannier-Frenkel ratio $(m_e+m_h)/M_F\equiv$wf. The Heller Marcus effect
shifts the mass away from negative values. As $\alpha^{-1} \rightarrow \infty$,
the curves become periodic in the inverse electric field $1/\alpha$, with identical period $P$ given by
$P=\frac{\pi}{4}(V_o-H_F)/t\approx 3.927$ .}
\label{fig:page73}
\end{figure}

\begin{figure}[htp]
\centering
\includegraphics[width=8.5cm,height=7cm]{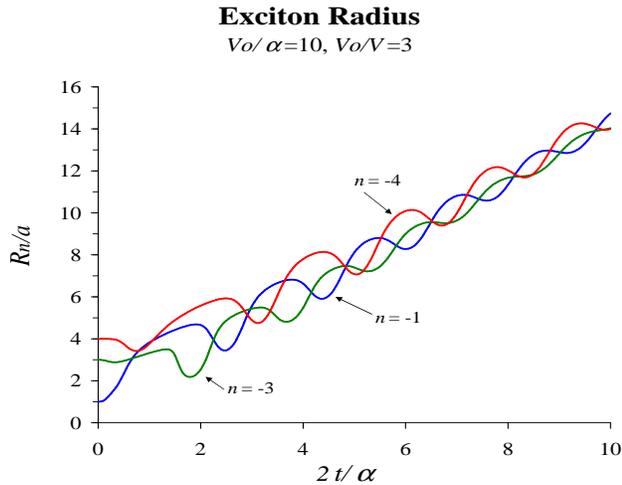}
\caption{Plot of the dimensionless exciton radius $R_n/a$ vs $2t/\alpha$
for three levels ($n=-1,-3,-4$), with $V_o/\alpha=10$ and $V_o/V=3$.
At $t = 0$ exciton motion is frozen, with expected
inter-particle distance equal to $|n|$. As $t \rightarrow \infty$
exciton radius diverges in an oscillatory manner.}
\label{fig:c3g23}
\end{figure}

\begin{figure}[htp]
\centering
\includegraphics[width=8.5cm,height=7cm]{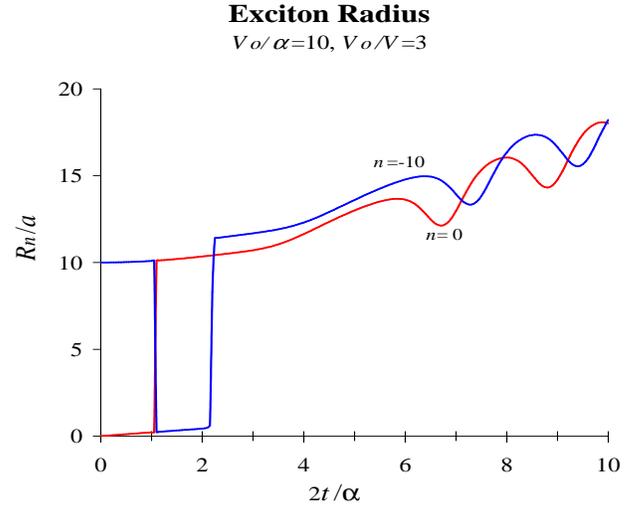}
\caption{Plot of the dimensionless exciton radius $R_n/a$ vs $2t/\alpha$ for two levels $n=0$ (red) and $n=-10$
(blue),
with $V_o/\alpha=10$ and $V_o/V=3$. The exciton radius shows abrupt variations due to the strongly
avoided crossings in the corresponding energy spectrum (figure \ref{fig:page20}). Note the envelope associated
to each one of the exciton radii, arising from the envelope in the Wannier-Stark ladder.}
\label{fig:c3g17}
\end{figure}

\begin{figure}[htp]
\centering
\includegraphics[width=8.5cm,height=7cm]{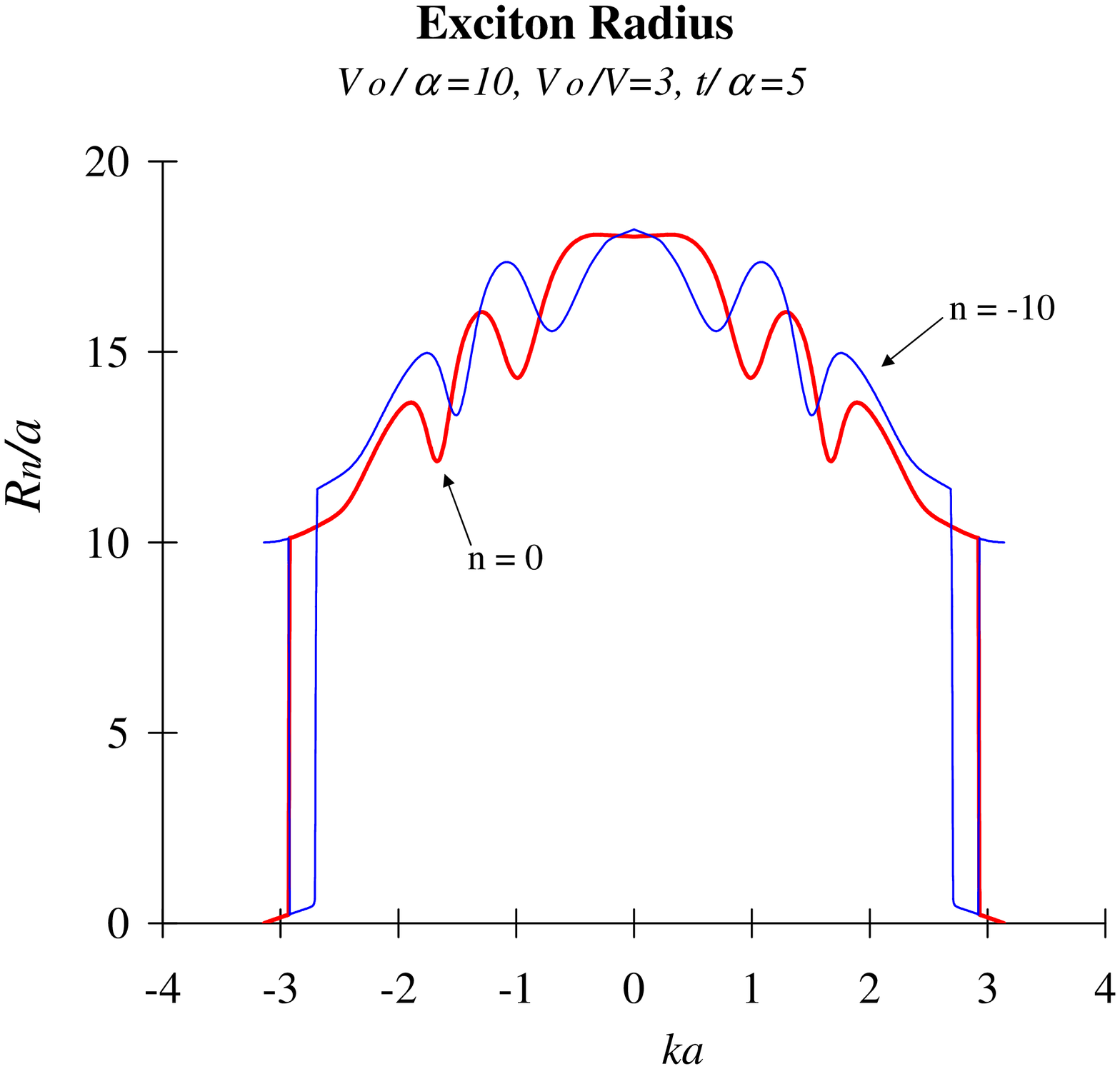}
\caption{Plot of the dimensionless exciton radii $R_n/a$ for $n=0$ (red)
and $n=-10$ (blue) vs $ka$ in the first Brillouin Zone. The strongly
avoided crossing of corresponding levels shown in figure
\ref{fig:page20} appears here as abrupt variations of the radii close to the
borders of the first Brillouin Zone.}
\label{fig:radio}
\end{figure}
%\includegraphics[scale=0.4]{radio}
%%%%%%%%%%%%%%%%%%%5

As for Fig (3), it was numerically obtained using a scaling
property \cite{MG-1984} (valid when $H_F=0$) related to the energy $E_n(k;t)$, namely,
\begin{equation}
E_n(k;t)-W=z_n(k;t)=z_n(0;t\cos{ka}).
\end{equation}

Different exciton masses are to be associated to levels $n=0$ and $n=-10$,
which have respective curvatures of different magnitude at $k=0$ in Fig (3).

Figures (4), (5) and (6) display a small representative sample of
our results for the effective exciton mass $M_n$ \cite{jean}.
Interestingly enough, we find both positive and negative effective
masses with $|\frac{m_e+m_h}{M_n}| \le 1$ when $M_F=\infty$ (i.e.,
in absence of the Heller-Marcus effect in Figures (4) and (5) ).
In Fig 6 we find that the exciton-hopping effect \cite{GM-1985}
shifts the mass away from negative values but still allows for
effective negative masses. That the effective mass of the exciton
may indeed become negative in the presence of an external electric
field is an interesting result found here that has not been
properly noted \cite{libroexciton} in the previous literature
\cite{MG-1984,Mattis,GM-1985,G-1996}, and might be of experimental
relevance for the excitonic Stark effect. The noticeable visual
envelope for the inverse mass in Fig (4) arises from the bound
state of Eq.(\ref{eq:boundstate}), or envelope profile of Fig (1).
It is given by \cite{13}

\begin{equation}
\left( \frac{m_e+m_h}{M_n} \right) _{env.} = \frac{x}{\sqrt{1+x^2}},
\label{eq:massenvelope}
\end{equation}
where $x\equiv W/V_o$.

In Figures (5) and (6), the inverse of the mass is seen to
oscillate (for $\alpha^{-1} \rightarrow \infty$) as a function of
$1/\alpha$. The period $P$ of these Keldysh-like oscillations can
be shown to be (see Appendix)
\begin{equation}
P=\frac{\pi}{4}\left(\frac{V_o}{t}\right).
\label{eq:period}
\end{equation}

Numerical results \cite{jean} furthermore show
Eq.(\ref{eq:period}) to be still valid when the Coulomb
interaction $V$ is present, i.e., the period of these oscillations
is independent of $V$.

Our three remaining figures display typical behavior of the
root-mean-square radius $R_n$ of the exciton in our model. Figures
(7) and (8) display an oscillatory behavior as $R_n$ diverges with
increasing $t \rightarrow \infty$. Superimposed on both figures
(7) and (8) is an overall hyperbolic ``pattern'', given by the
formula (see Appendix)
\begin{equation}
\frac{R_n}{a}=\sqrt{\frac{8\,t^2}{\alpha^2}+n^2},
\label{eq:radiusa}
\end{equation}
which expresses the value of $R_n$ under the action of the
electric field alone, i.e., for a homogenous equally spaced
Wannier-Stark ladder, with eigenvalues $E_n-W=n\,\alpha$.

In Fig.(8) the radius $R_n/a$ for $n=0$ starts at zero, and for
small enough $2\,t/\alpha$ it increases linearly (cf.
Eq.(\ref{eq:radiusa})) until it reaches the value of $2\,t/\alpha
\sim 1$. At this point one observes an abrupt variation of
$R_n/\alpha$ for both levels (n=0, -10), and a simultaneous
strongly avoided crossing in the Wannier-Stark ladder. From then
on the contribution to the linear increase in the radius (cf.
Fig.(2)) is given by the following level down in the Wannier-Stark
ladder (here n=-10). These abrupt variations in the exciton's
size, exemplified here with these two levels ($n=0, -10$), are,
however, a general feature of our model \cite{jean}. One which
will always occur associated to a strongly avoided crossing in the
respective eigenvalue spectrum, and one which has no counterpart
in the study of, for example, the standard Stark effect for the
hydrogen atom \cite{callaway}.

Finally, in Fig. (9) we display the $k$-dependence of $R_n/a$
for the same levels as in Fig.(8) at $t/\alpha=5$. The abrupt
variations in Fig. (8) are displayed in Fig. (9) close to the
first Brillouin Zone borders, and oscillations of $R_n/a$
in one curve are transposed onto the other.

Having displayed graphically some of our extensive \cite{jean}
numerical results, in the next section some conclusions are
presented and comments made about the relevance of our exciton
model.

\section{\label{sec:conclusions}Conclusions}
Although simple, the exciton model considered previously displays
quite interesting behavior. Most notably, oscillating (negative)
excitonic masses and abrupt variations in the exciton's size
(among others). To the best of our knowledge, these are novel
predictions that have not been considered before in the
literature, nor have these effects been experimentally observed.
Since the strongly ionizing influence of an external
electric field upon a weakly bound exciton is
well-known \cite{libroexciton}, it is clear that the difficulty
of experimental observation of some of these effects might be a large one.
Furthermore, it is unclear to what extent the previous theoretical
results may be modified by, for example, the inclusion of a more
realistic exciton (band) dispersion law, as well as the lifting of
the one-dimensionality of the model. To this extent, further
investigation of our model's interesting predictions is warranted
in the future.

From a self-consistent point of view it is worthwhile to point out
that, within the context of the model's given tight-binding
dispersion, the semiclassical effective potential \cite{Ashcroft}
$V_{eff.}(n)$ associated to $V(n)$ (cf. Eq. ({\ref{eq:ec-2}})),
is given by \cite{unpublished}
\begin{equation}
V_{eff.}(n)=\mu^*\left[\frac{a^2}{2}\,V^{2}(n) - a^2\, \verb"E" \, V(n)\right],
\end{equation}
(where $ \verb"E" $ is the semiclassical energy of the exciton and
$\mu^*$ is the reduced mass of the electron and hole).
Since $V_{eff.}(n)$ is positive and diverges for $|n|\rightarrow \infty$,
this precludes the occurrence of the well-known \cite{libroexciton}
tunneling through the potential barrier that appears in the standard Stark
effect for the hydrogen atom, or for a non tight-binding exciton in the
parabolic effective mass approximation \cite{libroexciton}.
Thus, our bound exciton states on the
lattice are truly localized \cite{ruso}, instead of being simply metastable
ones as for a hydrogen-like atom on the continuous manifold
\footnote{Furthermore, our Green's function matrix elements $G_{m,m}(z)$
have no cuts, i.e., they display only isolated poles on the real axis.}, and
consequently, both the mass and the radius of the exciton are well-defined
concepts for these states. It is to be noticed, finally, that although presented
in the context of the (MGF), our exact results for the effective mass $M_n$ of
the exciton have also been obtained independently \cite{jean} from the exciton
dispersion law $E_n(k)$. Thus the validity of the (MGF) has in this respect
also been checked.

In summary, we have considered a very simple one-dimensional exciton model in
the presence of a static and homogenous electric field, and this model has been
found to display highly interesting behavior for some of its properties, such
as mass and radius \footnote{Largely independent of boundary conditions, since
it is noteworthy that placing our system in a finite ``box" (so as to make the limit
$\alpha \rightarrow 0$ non-singular) does not alter significantly any of
our physical results \cite{jean}.}. Further theoretical work should then be able to elucidate
to what extent some of these predicted effects might be experimentally observable.

\appendix*

\section{}

In this Appendix we derive several analytical results used in this
paper. First, we obtain Eq.(\ref{eq:radius}) in the text from the
definition in Eq.(\ref{eq:radiusf}). Based on the Green's function
formalism \cite{economou}, the continued fraction $G_{m,m}(z)$ can
be written as:
\begin{equation}
G_{m,m}(z)=\sum_{m'}\frac{|F_{m'}(m)|^2}{z-z_{m'}}.
\label{eq:green}
\end{equation}
where the $z_{m'}$'s are simple poles of $G_{m,m}(z)$. Multiplying
both sides of Eq. (\ref{eq:green}) by $(z-z_{n})$, we obtain:
\begin{equation}
\left(z - z_{n}\right)
G_{m,m}(z)=\sum_{m'}\frac{z-z_{n}}{z-z_{m'}}|F_{m'}(m)|^2,
\end{equation}
and by taking the limit $(z\rightarrow z_{n})$, we get
\begin{equation}
\lim_{z\rightarrow z_{n}}\left(
z-z_{n}\right)G_{m,m}(z)=|F_{n}(m)|^2,
\end{equation}
or equivalently
\begin{equation}
\lim_{z\rightarrow z_{n}}\frac{\left(
z-z_{n}\right)}{G_{m,m}^{-1}(z)}
=\frac{1}{\frac{\partial}{\partial
z}\left(G_{m,m}^{-1}(z)\right)_{z=z_n}}=|F_{n}(m)|^2.
\label{eq:limitf}
\end{equation}
By substituting the result of Eq.(\ref{eq:limitf}) into
Eq.(\ref{eq:radiusf}) we obtain Eq.(\ref{eq:radius}).

Now we obtain Eq.(\ref{eq:period}) which gives the period $P$ of
the Keldysh-like oscillations observed in the energy and the
inverse mass of the exciton in the limit $t/\alpha \rightarrow
\infty$. On one hand, we start with Eq.(\ref{eq:frac}) written in
the absence of the Coulomb interaction, i.e., with $V(m)=-V_o
\delta_{m,0}+\alpha\,\, m$ \footnote{When there is
exciton-hopping, $V_o$ is simply replaced by $V_o-H_F$}, and then
we use the following relationship
\begin{equation}
\ds\frac{y}{2 \nu}\ds\frac{J_{\nu-1}(y)}{J_{\nu}(y)}-1=\fsm{\ds-\frac{\left(\ds\frac{y}{2}\right)^2}
{\nu(\nu+1)}}{1}{\ds\frac{\left(\ds\frac{y}{2}\right)^2}{(\nu+1)(\nu+2)}}{1}
{\ds\frac{\left(\ds\frac{y}{2}\right)^2}{(\nu+2)(\nu+3)}}{1},
\label{eq:fracbessel}
\end{equation}
which expresses Bessel functions $J_{\nu}(y)$ of the first kind
in a continued fraction form \cite{Abramowitz}. Using the following identities
\begin{eqnarray}
&&J_{\nu+1}(y) J_{-\nu}(y)+J_\nu(y)J_{-\nu-1}(y)=\frac{-2 \sin(\nu \pi)}{\pi y},
\nonumber
\\ \nonumber \\ &&J_{\nu-1}(y)+J_{\nu+1}(y)=\frac{2 \nu}{y} J_\nu(y),\hskip0.4cm \mbox{and}
\\ \nonumber \\ &&J_\nu(-y)=(-1)^\nu J_\nu(y),\label{eq:bessel}
\end{eqnarray}
Eq. (\ref{eq:frac}) can then be written (for $m=0$ and $\varepsilon \equiv 2t$) as
\begin{equation}
G^{-1}_{0,0}(z)=\frac{\alpha}{\pi}\frac{sen(\nu \pi)}{J_{\nu}(y) J_{-\nu}(y)}-V_o,
\label{eq:Ginverse}
\end{equation}
where $\nu \equiv z/\alpha$ and $y\equiv 4 t/\alpha$.

The eigenvalues $E_n$ are obtained from the following implicit equation upon the $z_n$'s, namely
\begin{equation}
\sin\left(\frac{z_{n}}{\alpha}\pi\right)=\left(\frac{V_o \pi}{\alpha}\right)J_{\frac{z_{n}}{\alpha}}
\left(\frac{4t}{\alpha}\right)J_{\frac{-z_{n}}{\alpha}}\left(\frac{4t}{\alpha}\right),
\label{eq:implicit}
\end{equation}
obtained in turn from $G^{-1}_{0,0}(z)=0$. When $\alpha^{-1}
\rightarrow \infty$, we use the asymptotic Bessel functions form
\cite{Abramowitz}
\begin{equation}
\lim_{y\rightarrow \infty} J_{\pm \nu}(y)=\sqrt{\frac{2}{\pi y}}\cos\left( y\mp \frac{1}{2}\nu
\pi-\frac{1}{4}\pi\right).
\label{eq:asymptotic}
\end{equation}
\\
Upon substitution of Eq.(\ref{eq:asymptotic}) into
Eq.(\ref{eq:implicit}), we obtain after some algebraic
manipulations that
\begin{eqnarray}
\frac{4t}{V_o}\sin\left( \frac{z_n}{\alpha}\pi\right) &=&\cos^2\left( \frac{1}{2}
\frac{z_n}{\alpha}\pi\right)-\sin^2\left( \frac{1}{2}\frac{z_n}{\alpha}\pi\right)
\nonumber \\
\nonumber \\ &&+\,2 \,\sin \left( \frac{4t}{\alpha}\right)\cos\left(
\frac{4t}{\alpha}\right), \;\,\mbox{or}  \nonumber \\
\nonumber \\
\frac{4}{A}\sin\left( \frac{z_n}{\alpha}\pi\right) &=&\cos\left( \frac{z_n}{\alpha} \pi\right)+
\sin\left(\frac{8}{A}\frac{V_o}{\alpha}\right);
\end{eqnarray}
${z_n}/{\alpha}$ as a function of $V_o/\alpha$ will then be periodic with period
$P=\frac{\pi}{4} A$, where $A\equiv\frac{V_o}{t}$, and so will the inverse mass of
the exciton according to Eq.(\ref{eq:MGF}).

To end this Appendix we indicate briefly the derivation of
Eq.(\ref{eq:radiusa}). By setting $V_o=V=H_F=0$ in Eq.
(\ref{eq:frac}), we find
\begin{equation}
 G^{-1}_{m,m}(z)=\frac{(-1)^m \ds\frac{\alpha}{\pi}\sin\left(
\frac{z} {\alpha}\pi\right)}{\ds J_{\frac{z}{\alpha}-m}\left(
\frac{4t}{\alpha}\right) \ds J_{-\frac{z}{\alpha}+m}\left(
\frac{4t}{\alpha}\right) }, \label{eq:gmm} \end{equation}
which can be proven \cite{jean} in a manner analogous to that of Eq.
(\ref{eq:Ginverse}). One now evaluates the derivative of $ G^{-1}_{m,m}(z)$ appearing
in Eq. (\ref{eq:radius}), as
\begin{equation}
\frac{\partial}{\partial z}G^{-1}_{m,m}(z)|_{z=z_n}=\frac{(-1)^{n+m}}{J_{n-m}\left(
\frac{4t}{\alpha} \right)  J_{-n+m}\left( \frac{4t}{\alpha}\right)},
\label{eq:A10}
\end{equation}
where $z_n=n\,\alpha$. Substituting Eq. (\ref{eq:A10}) into Eq. (\ref{eq:radius}),
there results
\begin{equation}
R_n^2=\sum_{m=-\infty}^{m=+\infty} (-1)^{m+n}\,a^2\,m^2\, J_{(n-m)}\left(
\frac{4t}{\alpha} \right)   J_{-(n-m)}\left( \frac{4t}{\alpha}\right),
\end{equation}
which reduces to
\begin{equation}
\frac{R^2_n}{a^2}=\sum_{m=-\infty}^{m=+\infty}\,m^2 J^2_{n+m}\left( \frac{4t}{\alpha} \right),
\label{eq:eqA15}
\end{equation}
by using Eq. (\ref{eq:bessel}). Eq. (\ref{eq:radiusa}) finally ensues
after some algebraic manipulations \cite{jean} on Eq. (\ref{eq:eqA15})

\bibliography{exciton}% Produces the bibliography via BibTeX.

\end{document}